\documentclass[showpacs,amsmath,amssymb,pre,twocolumn]{revtex4}

\usepackage{graphicx}

\newcommand*{\dd}{\mathrm{d}}
\newcommand*{\ee}{\mathrm{e}}
\newcommand*{\iu}{\mathrm{i}}
\newcommand*{\ind}[1]{_{\text{#1}}}
\newcommand*{\hoch}[1]{^{\text{#1}}}
\newcommand*{\ket}[1]{\mathinner{\lvert#1\rangle}}
\newcommand*{\bra}[1]{\mathinner{\langle#1\rvert}}
\newcommand*{\ketbra}[2]{\ket{#1}\!\bra{#2}}
\newcommand*{\expec}[1]{\mathinner{\langle#1\rangle}}
\newcommand*{\abs}[1]{\lvert#1\rvert}
\newcommand*{\op}{\hat}
\newcommand*{\sop}{\op\sigma}

\newcommand*{\Tr}{\operatorname{Tr}}
\newcommand*{\mat}{\mathsf}

\newcommand*{\epsinv}{\lambda_\varepsilon}

\newcommand*{\cs}{S}
\newcommand*{\env}{C}

\newcommand*{\Hoff}{\op{H}\ind{off}}
\newcommand*{\Hdiag}{\op{H}\ind{diag}}

\begin{document}

%\title{Canonical and Microcanonical Contraints}

\title{Non Thermal Equilibrium States of Closed Bipartite Systems}

\author{Harry Schmidt} %
\email[E-Mail: ]{harry.schmidt@itp1.uni-stuttgart.de}
\affiliation{Institut f\"ur
  Theoretische Physik 1, Universit\"at Stuttgart}

\author{G\"unter Mahler} 
\affiliation{Institut f\"ur
  Theoretische Physik 1, Universit\"at Stuttgart}

\begin{abstract}
  We investigate a two-level system in resonant contact with a
  larger environment.  The environment typically is in a canonical
  state with a given temperature initially.  Depending on the precise
  spectral structure of the environment and the type of coupling
  between both systems, the smaller part may relax to a canonical
  state with the same temperature as the environment (i.e.~thermal
  relaxation) or to some other quasi equilibrium state (non thermal
  relaxation).  The type of the (quasi) equilibrium state can be
  related to the distribution of certain properties of the energy
  eigenvectors of the total system.  We examine these distributions
  for several abstract and concrete (spin environment) Hamiltonian
  systems, the significant aspect of these distributions can be
  related to the relative strength of local and interaction parts of
  the Hamiltonian.
\end{abstract}


\maketitle

\section{Introduction}
\label{sec:intro}

In a composite but closed quantum system in which a smaller central
system~\cs\ is weakly coupled to a larger environment~\env, most of
the (pure) states of the total system for a given energy (and possibly
some additional constraints) exhibit properties of thermal equilibrium
states with respect to the smaller part~\cite{mahler:quant_therm},
i.e.~there exists a so-called dominant region in Hilbert space in
which the entropy of the central system is close to its maximum value
under the given constraints.  Therefore for most pure initial states
of the total system, the state of the central system shows decoherence
and some kind of thermalization; it typically approaches a
quasi-equilibrium canonical state with a temperature given by the
spectral properties of the environment~\cite{borowski.2003}.

If the environment initially is in a thermal state with a given
temperature and consists of many bands or of a broad continuum of
levels, the central system typically relaxes to a thermal state with
the same temperature.  This type of relaxation will be called
\emph{thermal} relaxation in the following.  Here we investigate to
what extent certain structures of the total system influence the
reached (quasi) equilibrium state.  We will relate this equilibrium
state to the distribution of the energy eigenvectors of the system, or
rather certain important aspects of this distribution.  We will show
that there is a close relation between the two, and how this affects
the equilibrium state for different system structures.

We particularly focus on a single spin-$1/2$ particle coupled to an
environment of spin-$1/2$ particles.  Recently, the properties of spin
systems of different structure (rings, stars, and others) have been
subject of extensive interest.  A lot of work has been done on the
question of
entanglement~\cite{briegel.2001,oconnor.2001,wang.2002,hutton.2004,hutton.2004a,vidal.2003},
their relaxation behavior has been
addressed~\cite{breuer.2004,lages.2004} and various techniques were
suggested to make any spin interact with any other
spin~\cite{imamoglu.1999,makhlin.1999}.

Here we extend our analysis form our previous
paper~\cite{schmidt.2005} regarding the controllability of relaxation
behavior within these spin systems

\section{Canonical and Non Canonical Relaxation}
\label{sec:canon-nonc-relax}

\begin{figure}
  \centering
  \includegraphics{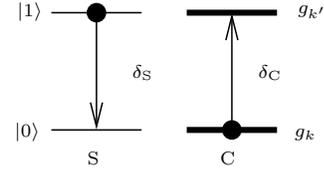}%
  \caption{A two-level system~\cs\ in contact with an
    environment~\env\ consisting of two highly degenerate
    levels~$k$,~$k'$ with degeneracies~$g_k$,~$g_{k'}$.}
  \label{fig:system}
\end{figure}

Figure~\ref{fig:system} shows a two-level system~(TLS) in
resonant~($\delta\ind{\cs}=\delta\ind{\env}=\delta$) contact with an
environment consisting of two ``energy bands''~$k$,~$k'$ of
degeneracies~$g_k$,~$g_{k'}$, respectively (for simplicity we
use~$g=g_k$, $g'=g_{k'}$ in the following, typically~$g'>g$).  The
coupling is assumed to be weak, the total system is described by the
Hamiltonian
\begin{equation*}
  \op{H} = \op{H}\ind{\cs} + \op{H}\ind{\env} + \op{H}\ind{int}.
\end{equation*}
A non equilibrium
state~$\ket{1}\ind{\cs}\otimes\ket{\phi_k}\ind{\env}$ is depicted
(here $\ket{\phi_k}$ denotes an arbitrary pure environmental state in
band~$k$).

If this state is taken as the initial state of a Schr\"odinger time
evolution of the total system, a relaxation to an equilibrium
situation is expected in which the time-averaged reduced state
operator of~\cs\ is given by~\cite{mahler:quant_therm}
\begin{equation}
  \label{eq:1}
  \op\rho\ind{\cs} = \frac{1}{g+g'} \bigl( g' \ketbra{0}{0} + g
  \ketbra{1}{1} \bigr)
\end{equation}
which can be interpreted as a canonical state operator with inverse
temperature
\begin{equation*}
  \beta\ind{\cs} = \frac{1}{k\ind{B}T} = 
  \frac{1}{\delta\ind{\cs}} \ln \frac{g'}{g}.
\end{equation*}
For a finite environment and weak random coupling, the reduced state
of the central system after relaxation still fluctuates
around~\eqref{eq:1}, see~\cite{borowski.2003}.

If the central system relaxes to state~(\ref{eq:1}) for any pure
state~$\ket{\phi_k}$ of the environment in band~$k$, it will relax to
the same state for an initial state in which the environment is
completely mixed within band~$k$,
$\op\rho_0=\ketbra{1}{1}\otimes\frac{\op{1}_k}{g}$.
$\op{1}_k$~denotes the projector onto band~$k$ of the environment.

Assume now that the environment is given by a large number $N$~of two
level systems with a homogeneous Zeeman
splitting,~$\op{H}\ind{\env}=\sum_{i=1}^N\delta\sop_z^{(i)}$.  The
environment initially is taken to be in a thermal state of
temperature~$\beta\ind{\env}$,
\begin{equation*}
  \op\rho_0 = \ketbra{1}{1} \otimes
  \frac{1}{Z}\, \ee^{-\beta\ind{\env}\op H\ind{\env}}.
\end{equation*}
If each band of the environment separately leads to a relaxation into
a state~\eqref{eq:1}, the equilibrium state of the central system is
given by the canonical state with~$\beta\ind{S}=\beta\ind{C}$ for
large~$N$.  For finite~$N$, the population of the excited state
of~\cs\ after relaxation is given by
\begin{equation*}
  (\op\rho\ind{\cs})_{11} = \frac{N}{N+1}\frac{1}{1+\ee^{\delta\beta\ind{\env}}} +
  \frac{1}{N+1}\; \xrightarrow[N\gg 1]{}\;
  \frac{1}{1+\ee^{\delta\beta\ind{\env}}}.
\end{equation*}
Because of this we call the relaxation from an initial
state~$\ket{1}\ind{\cs}\otimes\ket{\phi_k}\ind{\env}$ to the reduced
equilibrium state~\eqref{eq:1} \emph{canonical} or \emph{thermal}
throughout this text.

\begin{figure}
  \centering
  \includegraphics{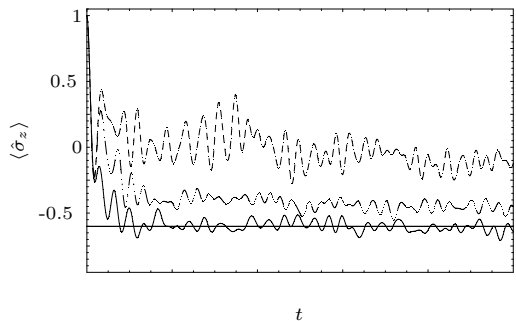}%
  \caption{$z$-component of the Bloch vector of~\cs\ for the initial
    state depicted in figure~\ref{fig:system}.  The environment
    consists of spins, curves are for different types of
    intra-environmental coupling. Fluctuations are due to the finite
    size of the system.  The black line indicates canonical
    equilibrium.}
  \label{fig:z-comp-different-couplings}
\end{figure}

However, not all environments lead to canonical relaxation of the
central system.  In~\cite{schmidt.2005} we examined several types of
spin environments and showed that many of these lead to an equilibrium
state that differs from~\eqref{eq:1}.  We call these deviations non
canonical or non thermal.  Figure~\ref{fig:z-comp-different-couplings}
shows the relaxation behavior for different types of environments.
Obviously, not all relax to the same quasi equilibrium state.   

\section{Energy Eigenvector Distributions}
\label{sec:eigenv-distr}

We now correlate deviations from canonical relaxation with the
distribution of the energy eigenvectors of the total system.  We only
consider the situation depicted in figure~\ref{fig:system}, since
environments with more bands in a canonical state can simply be
derived from these results.  As long as the interaction is weak
relative to the band splitting, the total system can be reduced to the
subspace consisting of the ``cross states''
\begin{gather}
  \{{\ket{0\ind{\cs},\text{environment in band }k'}},\quad  \notag\\
  \quad\ket{1\ind{\cs},\text{environment in band }k}\}.
  \label{eq:2}
\end{gather}
In the following we will always refer only to this
$(g+g')$-dimensional subspace.

Instead of the temperature we consider the population
inversion~$\Tr\{\sop_z\op\rho\ind{\cs}\}$ of the central system for
mathematical convenience.  The inversion of the canonical
state~\eqref{eq:1} is given by
\begin{equation*}
  \expec{\sop_z}\ind{can} = \frac{g-g'}{g+g'}.
\end{equation*}

The energy eigenvectors can be written in the form
\begin{equation}
  \label{eq:3}
  \ket{\varepsilon} \approx \alpha_\varepsilon \ket{0,\chi_\varepsilon} +
  \beta_\varepsilon \ket{1,\eta_\varepsilon},
\end{equation}
where~$\ket{\chi_\varepsilon}$ is a state in band~$k'$,
and~$\ket{\eta_\varepsilon}$ a state in band~$k$ of the environment.
In the following we will use~$\varepsilon$ as a discrete index running
from~$1$ to~$g+g'$ to number the energy eigenvectors within the
subspace of Hilbert space spanned by the states~\eqref{eq:2}.

We expand the initial
state~$\op\rho_0=\ketbra{1}{1}\otimes\frac{\op{1}_k}{g}$ in terms of
these eigenvectors,
$\op\rho_0=\sum_{\varepsilon,\varepsilon'}\rho_{0,\varepsilon\varepsilon'}\ketbra{\varepsilon}{\varepsilon'}$.
Averaging over all times and tracing out the environment yields the
equilibrium state of the central system,
\begin{align*}
  \bar\rho\ind{\cs} &= \sum_\varepsilon\rho_{0,\varepsilon\varepsilon} 
  \Tr\ind{\env}\bigl\{\ketbra{\varepsilon}{\varepsilon}\bigr\}\\
  &=
  \sum_\varepsilon \abs{\beta_\varepsilon}^2 \bigl(
  \abs{\alpha_\varepsilon}^2 \ketbra{0}{0} +
  \abs{\beta_\varepsilon}^2 \ketbra{1}{1} \bigr).
\end{align*}
The respective inversion is
\begin{equation*}
  \expec{\bar\sigma_z} = \sum_\varepsilon \abs{\beta_\varepsilon}^2
  \bigl( 
%  \underbrace{
    \abs{\beta_\varepsilon}^2 -
    \abs{\alpha_\varepsilon}^2
  %}_{\epsinv \text{ (see text)}}
  \bigr).
\end{equation*}
We notice
that
\begin{equation}
  \epsinv = \abs{\beta_\varepsilon}^2-\abs{\alpha_\varepsilon}^2
\end{equation}
is the inversion of~$\ket{\varepsilon}$.
Since~$\abs{\beta_\varepsilon}^2+\abs{\alpha_\varepsilon}^2=1$ we can
rewrite the time-averaged inversion of the central system completely
in terms of the~$\epsinv$'s,
\begin{equation*}
  \expec{\bar\sigma_z} = \frac{1}{2}\biggl(
  \sum_\varepsilon \epsinv + \sum_\varepsilon \epsinv^2 \biggr).
\end{equation*}
The first sum can be shown to be equal to~$g-g'$, see
appendix~\ref{sec:sum-lambda}.  If we rewrite
the second sum in terms of the variance~$\Delta\epsinv^2$ of the
$\epsinv$-distribution, we finally get
\begin{equation}
  \label{eq:4}
  \expec{\bar\sigma_z} = \expec{\sop_z}\ind{can} +
  \frac{(g+g')}{2g}\, \Delta\epsinv^2.
\end{equation}
The average deviation from the canonical equilibrium state is thus
mainly given by the distribution of the reduced states of the energy
eigenvectors, in particular by its variance.

As long as the width of the distribution is finite, there is always a
deviation from the canonical inversion and therefore also from the
canonical temperature.  For a finite environment this is always the
case.  We will now consider several types of environments and
interactions.

\section{Eigenvectors Distribution for Different Hamiltonians}
\label{sec:eigenv-distr-diff}

\subsection{Random Hamiltonian}
\label{sec:random-hamiltonian}

\begin{figure}
  \centering
  \includegraphics{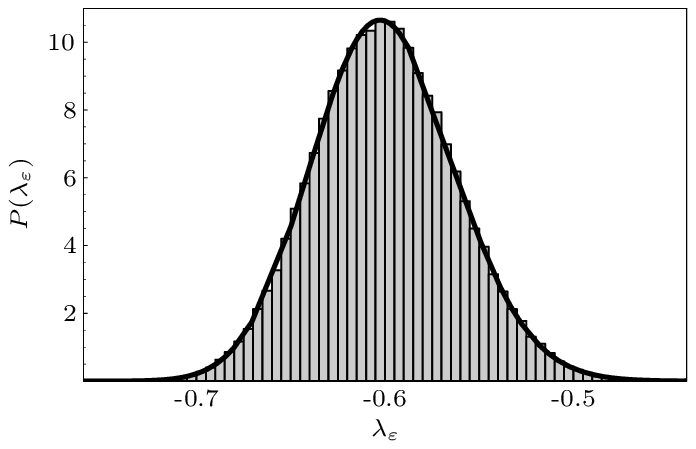}%
  \caption{$\epsinv$-distribution for random Hermitian matrices from
    the GUE.  The solid line shows the (normalized) probability
    density~(\ref{eq:5}), the histogram is calculated form the
    eigenvectors of 400 different random matrices. $g=91$, $g'=364$,
    $\Delta\epsinv^2=2/1425\approx0.0014$,
    $\expec{\sop_z}\ind{can}=-0.6$.}
  \label{fig:EvecRed-randomHamiltonian}
\end{figure}

At first we assume that the two relevant bands in the environment are
exactly degenerate and the transitions within~\cs\ and~\env\ are
exactly in resonance.  In this
case,~$\op{H}\ind{\cs}+\op{H}\ind{\env}\propto\op{1}$ (within the
relevant subspace spanned by the states~\eqref{eq:2}), thus we only
need to deal with~$\op{H}\ind{int}$.

If the coupling~$\op{H}\ind{int}$ between system and environment is
modeled by a random Hermitian matrix with a uniform Gaussian
distribution~$w(\op{H}\ind{int})=C\exp(-A\Tr\{\op{H}\ind{int}\}^2)$
(i.e.~taken from the GUE~\cite{haake:quant_chaos}), the state of the
central system typically relaxes to the expected equilibrium state
under Schr\"odinger dynamics for the total
system~\cite{mahler:quant_therm,borowski.2003,schmidt.2005}.  The
state fluctuates in time, the amplitude of these fluctuations
decreases with the size of the environment.

Figure~\ref{fig:EvecRed-randomHamiltonian} shows the
$\epsinv$-distribution for these random Hermitian matrices.  The
histogram was obtained by choosing a number of random matrices with
the given probability distribution and calculating the inversion of
their eigenvectors.  The parameters used are~$g=91$
and~$g'=364$.

The probability density for the~$\epsinv$'s for random Hermitian
matrices from the GUE can be calculated analytically and is given by
\begin{equation}
  \label{eq:5}
  P(\lambda) \propto (1-\lambda)^{g'-1} (1+\lambda)^{g-1}.
\end{equation}
This result is derived in appendix~\ref{sec:app-a}.  The solid line in
figure~\ref{fig:EvecRed-randomHamiltonian} shows the distribution.

The mean value of~$\epsinv$ for this distribution is given by
\begin{equation*}
  \bar\lambda_\varepsilon = \frac{g-g'}{g+g'} = \expec{\sop_z}\ind{can},
\end{equation*}
as expected from the general result in appendix~\ref{sec:sum-lambda}.
This is again the mean inversion of the equilibrium state for
canonical relaxation.

The variance of this distribution is given by
\begin{equation*}
  \Delta\epsinv^2 = 4 \frac{g\cdot g'}{d^2(d+1)}
\end{equation*}
with~$d=g+g'$.  Since this is finite, the reached steady state will
typically deviate from the canonical equilibrium state.

For large systems, i.e.~$d\rightarrow\infty$ (for constant
ratio~$g/g'$), both $\Delta\epsinv^2$
and~$\expec{\bar\sigma_z}-\expec{\sop_z}\ind{can}$ vanish.  So in the
thermodynamic limit, the quasi equilibrium reached equals canonical
equilibrium.

\subsection{Random interaction}
\label{sec:random-interaction}

\subsubsection{Degenerate bands}
\label{sec:degenerate-bands}

We will now discuss a system that is not completely random, but has a
random energy exchanging coupling between the central system and the
environment.  If the environmental bands are strictly degenerate, the
Hamiltonian matrix has the form
\begin{equation}
  \label{eq:6}
  \left(
  \begin{array}{@{}ccc|ccc@{}}
    -\varDelta/2     &         &        &   \\
            & \ddots &         &        & \mat{V}^\dagger\\
            &        & -\varDelta/2 \\ \hline
            &        &         & \varDelta/2 \\
            & \mat{V} &&& \ddots \\
            &        &         &        &        & \varDelta/2
  \end{array}
  \right),
\end{equation}
where~$\varDelta=\delta\hoch{\cs}-\delta\hoch{\env}$ is the detuning
between system and environment.  The upper left block corresponds to
the ground state of the central system, its therefore of
dimension~$g'$, the lower right block (of dimension~$g$) corresponds
to the excited level.  These blocks are purely diagonal.

The off diagonal block~$\mat{V}$ (a $g\times{}g'$~matrix),
corresponding to energy exchange (canonical) coupling, is chosen
randomly with normalized Gaussian distributions for the real and
imaginary parts of the matrix elements.

For~$\varDelta\neq0$, system and environment are off resonance and the
relaxation to the canonical equilibrium state is prohibited by energy
conservation.  So just~$\varDelta=0$ is considered.

\begin{figure}
  \centering
  \includegraphics{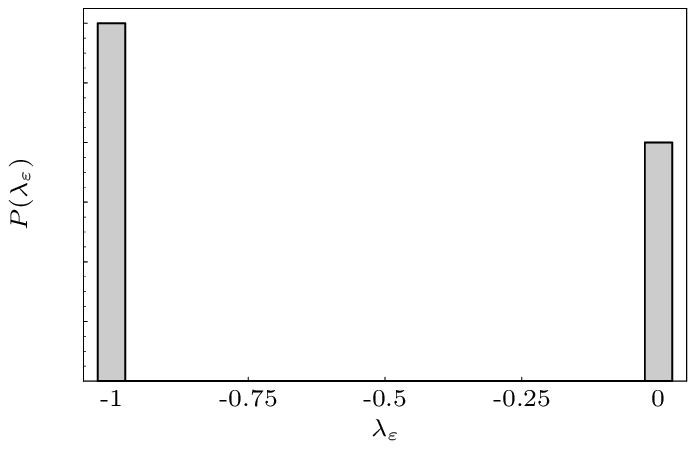}%
  \caption{Random interaction, degenerate bands.  The two bars
    represent delta-peaks of different strength.
    $\Delta\epsinv^2=6/25=0.24$.}
  \label{fig:EvecRed-randomInteraction-degen}
\end{figure}

Figure~\ref{fig:EvecRed-randomInteraction-degen} shows
the~$\epsinv$-distribution for a matrix of type~\eqref{eq:6}, again
for~$g=91$ and~$g'=364$.  Instead of being single peaked, the
distribution here is drastically different and consists of delta peaks
at~$\epsinv=-1$ and~$\epsinv=0$, respectively.  For a given
Hamiltonian, there are exactly~$g'-g$ eigenstates with~$\epsinv=-1$
and~$2g$ eigenstates with~$\epsinv=0$ (assuming~$g'>g$).  The mean
value still is~$\bar\lambda_\varepsilon=(g-g')/(g+g')$, as expected.

\begin{figure}
  \centering
  \includegraphics{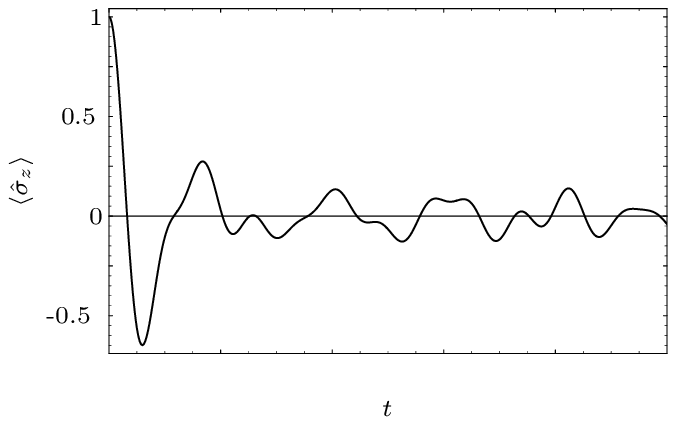}%
  \caption{Time evolution of the~$z$-component of the Bloch vector
    of~\cs\ for a random interaction and degenerate environmental
    bands, as described in section~\ref{sec:degenerate-bands}.  Note
    that the canonical equilibrium state for the given parameters
    would be at~$\expec{\sop_z}=-0.6$.}
  \label{fig:timeevol-degenerate}
\end{figure}

However, the variance of the distribution is obviously considerably
bigger.  For the given parameters~$\Delta\epsinv=0.24$ as opposed
to~$\Delta\epsinv\approx0.0014$ for the completely random Hamiltonian.
The deviation of the quasi equilibrium state for a Hamiltonian of
type~\eqref{eq:6} from the canonical one is exactly
\begin{equation*}
  \expec{\bar\sigma_z} - \expec{\sop_z}\ind{can} =
  -\expec{\sop_z}\ind{can}
\end{equation*}
which is nicely demonstrated in figure~\ref{fig:timeevol-degenerate}.
Since this result is independent of the system size, going to large
environments will not change the relaxation behavior other than
reducing the amplitude of the fluctuations.  Even in the thermodynamic
limit the canonical equilibrium state is never reached.

\subsubsection{Non degenerate bands}
\label{sec:non-degenerate-bands}

\begin{figure}
  \centering
  \includegraphics{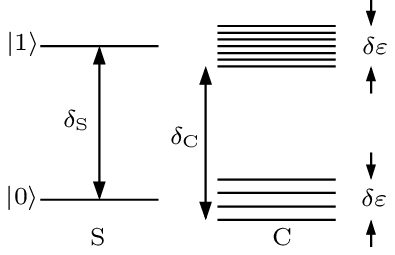}%
  \caption{A TLS in contact with an environment consisting of two
    ``bands'' with equidistant level spacing.  Both bands have the
    same width~$\delta\varepsilon$, the lower band consists of~$g$
    levels, the upper band of~$g'$ levels.}
  \label{fig:system_equidist}
\end{figure}

The situation, again, changes if we introduce a finite spacing between
the levels within each environmental band.  For simplicity we will
only consider equidistant levels and equal
bandwidth~$\delta\varepsilon$ for both bands here.  The lowest (and
highest) levels of each band in the environment are in resonance with
the central system.  This system is depicted in
figure~\ref{fig:system_equidist} ($\delta\ind{\cs}=\delta\ind{\env}$
is considered in the following).  The Hamiltonian matrix in
subspace~\eqref{eq:2} is given by.

\begin{equation*}
  \left(
    \begin{array}{@{}c|c@{}}
      \begin{array}{@{}cccc@{}}
        0 \\
        & \frac{\delta\varepsilon}{(g'-1)} \\
        && \ddots \\
        &&& \delta\varepsilon
      \end{array}
      & \mat{V}^\dagger \\ \hline
      \mat{V}^\dagger &
      \begin{array}{@{}cccc@{}}
        0 \\
        & \frac{\delta\varepsilon}{(g-1)} \\
        && \ddots \\
        &&& \delta\varepsilon    
      \end{array}
    \end{array}
  \right)
\end{equation*}

\begin{figure}
  \centering
  \includegraphics{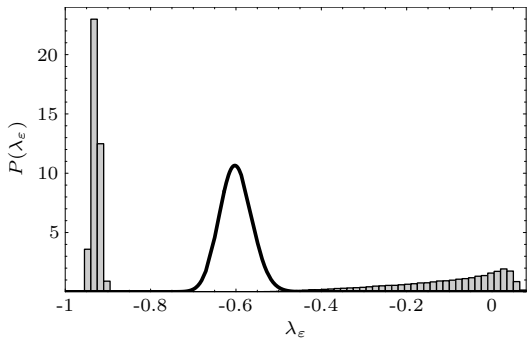}%
  \caption{Random interaction, equidistant spacing, small level
    spacing.  $\Delta\epsinv^2\approx0.19$.  The solid line shows the
    distribution for the complete random Hamiltonian.}
  \label{fig:EvecRed-randomInteraction-equidist-weak}
\end{figure}

By introducing a small level splitting in the environment (small
compared to the interaction strength), the peaks in
figure~\ref{fig:EvecRed-randomInteraction-degen} get broader,
especially the one at~$\epsinv=0$ gets flatter considerably and is
stretched towards negative~$\epsinv$.  The variance of the
distribution gets smaller.
Figure~\ref{fig:EvecRed-randomInteraction-equidist-weak} shows the
$\epsinv$-distribution for a relatively small level spacing.

\begin{figure}
  \centering
  \includegraphics{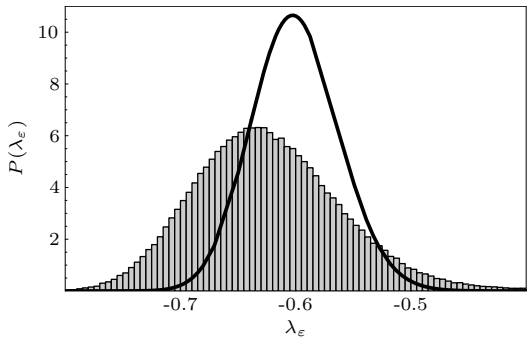}%
  \caption{Random interaction, equidistant spacing, larger level
    spacing.  The solid line shows the distribution for the complete
    random Hamiltonian. $\Delta\epsinv\approx0.022$.}
  \label{fig:EvecRed-randomInteraction-equidist}
\end{figure}

When the level splitting is increased, the distributions becomes
single peaked, with the peak close to its average and of similar
height as the peak of the complete random Hamiltonian, see
figure~\ref{fig:EvecRed-randomInteraction-equidist}.  A long tail
towards higher~$\epsinv$ prevails, therefore the variance is still
considerably larger.

\subsection{Spin environments}
\label{sec:spin-environments}

We now consider a gapped spin or TLS as in figure~\ref{fig:system}
coupled to an array of spins, all with a Zeeman splitting equal to the
central spin (or almost equal).  Due to energy conservation, the
system can be reduced to the situation shown in
figure~\ref{fig:system} for each pair of environmental bands.  If the
environment initially is in a canonical state, we can simply sum up
over all bands, as long as the interaction between the environmental
spins is small.

The $\epsinv$-distributions of several spin-environments have been
discussed in \cite{schmidt.2005}, so we will only discuss them briefly
here.

\subsubsection{Spin-star configuration}
\label{sec:spin-star-conf}

\begin{figure}
  \centering
  \includegraphics{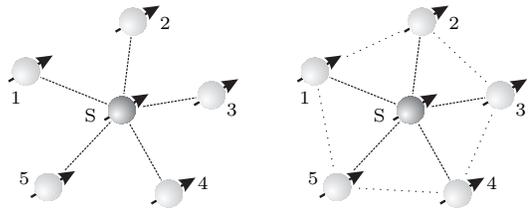}%
  \caption{Schematics of the spin-star (left) and spin-ring (right)
    configurations.  Typically, the environment consists of a lot more
    than five spins.}
  \label{fig:spin-star}
\end{figure}

Figure~\ref{fig:spin-star} (left) shows schematically a spin-star
configuration, i.e.~a central spin coupled to an array of
environmental spins without mutual interaction.  A typical environment
should of course consist of a lot more than 5~spins.  The most general
Hamiltonian describing the system-environment interaction is
\begin{equation*}
  \op{H}\ind{int} = \sum_{i,j=1}^3 \sum_{\nu=1}^N \gamma_{ij}^{(\nu)}
  \sop_i^{(\text{\cs})} \otimes \sop_j^{(\nu)}
\end{equation*}
for $N$~environmental spins.

\begin{figure}
  \centering
  \includegraphics{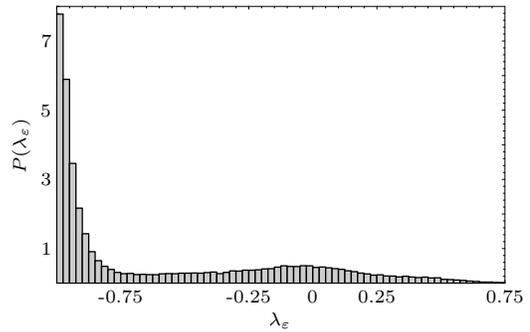}%
  \caption{$\epsinv$-distribution for the spin-star configuration.
    14~environmental spins, the 2nd and 3rd excited bands are
    considered, corresponding to $g=91$, $g'=364$.
    $\Delta\epsinv^2\approx0.216$.}
  \label{fig:EvecRed-spinstar}
\end{figure}

It has been shown in~\cite{schmidt.2005} that if the
coefficients~$\gamma_{ij}^{(\nu)}$ are chosen randomly, the initial
state depicted in figure~\ref{fig:EvecRed-randomInteraction-degen}
typically does not relax to the canonical equilibrium state.  The
Hamiltonian matrix in this case has the form~\eqref{eq:6}, however
with small fluctuations on the diagonal, and the interaction part of
the Hamiltonian matrix is only sparsely populated.  Nevertheless, the
$\epsinv$-distribution shows some  similarity to the one described in
section~\ref{sec:non-degenerate-bands} for small level splitting.
Figure~\ref{fig:EvecRed-spinstar} shows the $\epsinv$-distribution for
this system.

\subsubsection{Intra-environmental coupling}
\label{sec:spin-ring-conf}

If mutual coupling between the environmental spins is introduced, the
situation changes.  Figure~\ref{fig:spin-star} (right) schematically
shows next neighbor coupling in the environment (spin-ring
configuration), but other configurations are possible as well.  As
long as this coupling is weak, the system can still be considered band
wise.

\begin{figure}
  \centering
  \includegraphics{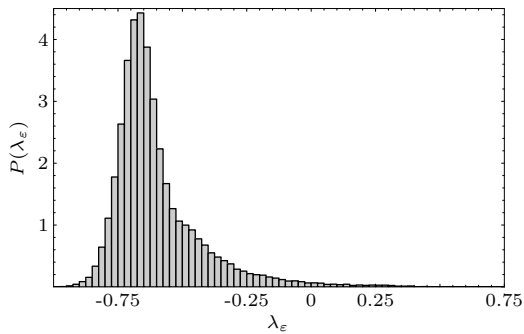}%
  \caption{$\epsinv$-distribution for a spin-ring configuration.
    14~environmental spins, the 2nd and 3rd excited bands are
    considered, corresponding to~$g=91$, $g'=364$.
    $\Delta\epsinv^2\approx0.0295$.}
  \label{fig:EvecRed-spinring}
\end{figure}

The interaction typically leads to a level splitting within the bands
which in turn leads to a $\epsinv$-distribution similar to the one
described in section~\ref{sec:non-degenerate-bands} with bigger level
splitting.  Figure~\ref{fig:EvecRed-spinring} shows the
$\epsinv$-distribution for a spin-ring configuration.  The shown
distribution is for a~$\sop_x\otimes\sop_x$ next neighbor coupling.
The distributions for different kinds of coupling, e.g.~Heisenberg
coupling, are similar.

\subsubsection{Inhomogeneous Zeeman splitting}
\label{sec:inhom-zeem-splitt}

\begin{figure}
  \centering
  \includegraphics{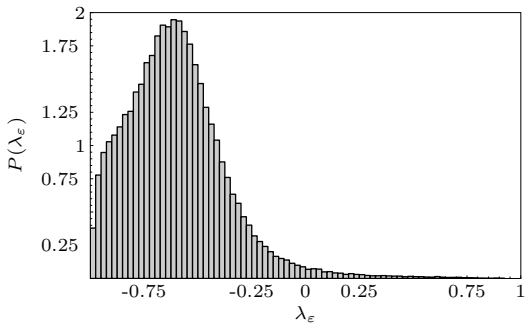}%
  \caption{$\epsinv$-distribution for a spin-star configuration with
    inhomogeneous Zeeman splitting of the environmental spins.
    14~environmental spins, the 2nd and 3rd excited bands are
    considered, corresponding to~$g=91$, $g'=364$.
    $\Delta\epsinv^2\approx0.0548$.}
  \label{fig:EvecRed-inhom-zee}
\end{figure}

If the individual environmental spins each have a different Zeeman
splitting, the situation becomes similar to the one discussed in
section~\ref{sec:non-degenerate-bands}.
Figure~\ref{fig:EvecRed-inhom-zee} shows the $\epsinv$-distribution
when the Zeeman splittings of the environmental spins are
homogeneously distributed within a certain range.  The distribution
again shows a peak around the mean
value~$\bar\lambda_\varepsilon=\expec{\sop_z}\ind{can}$, although
broader than in the previous case.

\section{Width of the distribution and spectral width}
\label{sec:width-distr-spectr}

Figures~\ref{fig:EvecRed-randomInteraction-degen},~\ref{fig:EvecRed-randomInteraction-equidist-weak},
and~\ref{fig:EvecRed-randomInteraction-equidist} indicate that there
is a continuous transition from a situation far from canonical to an
almost canonical relaxation, depending on the environmental spectrum.
What has been changed is the ``strength'' of the environmental
spectrum from zero to the minimal variance of the
$\epsinv$-distribution.  A similar transition can be observed for many
different environmental spectra.

In order to relate different types of spectra we split the Hamiltonian
matrix (in the considered subspace) in its respective diagonal and off
diagonal parts.  The off diagonal part~$\Hoff$ ($\mat{V}$
and~$\mat{V}^\dagger$ of~\eqref{eq:6}) describes the interaction
between central system and environment, while the diagonal
part~$\Hdiag$ describes the environmental spectra alone, if system and
environment are in resonance.  The diagonal part is always taken to be
traceless.  The quantity we use to compare different spectra is the
relative strength of the environmental part to the interaction part,
\begin{equation*}
  V\ind{R} = \sqrt{%
    \frac{\Tr(\op{H}\ind{diag}^2)}{\Tr(\op{H}\ind{off}^2)}.%
  }
\end{equation*}
For a completely random matrix (GUE) this relation is given on average
by
\begin{equation*}
  V\ind{R,GUE} = \sqrt{\frac{g^2+g'^2}{2gg'}}
\end{equation*}
which only depends on~$g/g'$, not on the actual size of the system.

\begin{figure}
  \centering
  \includegraphics{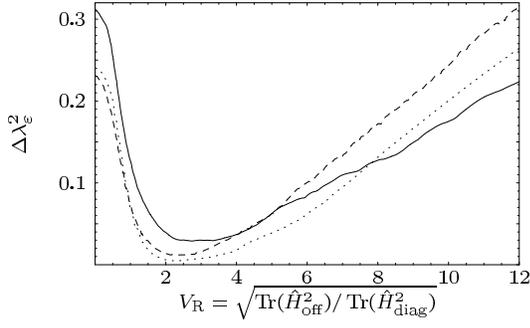}%
  \caption{Variance of the $\epsinv$-distribution over the relative
    strength of the environmental spectrum for different types of this
    spectrum.}
  \label{fig:variance-vs-relVar}
\end{figure}

Figure~\ref{fig:variance-vs-relVar} shows the variance of the
$\epsinv$-distribution for three different types of environmental
spectra.  In all three cases the environment consists of 14~spins, the
2nd and 3rd excited bands are considered, $g=91$,~$g'=364$, as
described in section~\ref{sec:spin-environments}.

The solid line corresponds to the spin-ring configuration as described
in section~\ref{sec:spin-ring-conf}. The intra-environmental
interaction is taken as a~$\sop_x\otimes\sop_x+\sop_y\otimes\sop_y$
next neighbor coupling, the central system is randomly coupled to each
environmental spin.  The dashed line corresponds to the configuration
described in section~\ref{sec:inhom-zeem-splitt}, there is no mutual
interaction between the spins in the environment, but their Zeeman
splitting is inhomogeneous.  The central system is again coupled
randomly to each environmental spin.  The environmental spectrum for
the dotted line is the same as for the dashed line. However, the
interaction between the central spin and each environmental spin is
modeled by~$\sop_x\otimes\sop_x+\sop_y\otimes\sop_y$.  The
corresponding average values for a random matrix from the GUE
are~$\Delta\epsinv^2=2/1425\approx0.0014$
and~$V\ind{R,GUE}\approx1.46$, respectively.

We notice that for each type of coupling and environmental spectrum
there is a distinct minimum of~$\Delta\epsinv^2$ for similar values
of~$V\ind{R}$ close to, but not exactly at the average value for the
GUE matrices.

\section{Conclusion}
\label{sec:conclusion}

We have characterized situations under which non-thermal states should
result as quasi-equilibrium states.  For a spin-$1/2$ particle weakly
coupled to a larger environment, there is a close relation between the
quasi-equilibrium state of the small quantum system coupled to a
larger environment and the distribution of certain properties of the
energy eigenvectors of the total system.  The equilibrium state is
directly given by the width of this distribution.  The spectral
structure of the environment and the exact form of the coupling has a
strong influence on the eigenvector distribution.  To show this we
have considered both abstract system Hamiltonians as well as
Hamiltonians for structured spin environments.

Furthermore, there is a close relation between the width of
the~$\epsinv$-distribution and the strengths of both the local
(diagonal) and the interaction (off diagonal) part of the Hamiltonian.
By changing certain parameters within each system, there is a distinct
minimum of the~$\epsinv$-width for a value of the relative strength
that's close to the respective value for GUE matrices.  This relative
strength can thus give an indication whether for a given system
relaxation to or close to a thermal state can be expected without
calculating the full~$\epsinv$-distribution.  The relative width gives
an indication how to choose the system parameters properly to achieve
a certain type of equilibrium situation.  This should be of help when
designing a spin system for special (``non-thermal'') relaxation
behavior.

\bigskip

We thank the Deutsche Forschungsgemeinschaft for financial support.

\appendix

\section{Towards eq.~(\ref{eq:4})}
\label{sec:sum-lambda}

Here we show that~$\sum_\varepsilon\epsinv=g-g'$.
\begin{align*}
  \sum_\varepsilon\epsinv &= \sum_\varepsilon
  \Tr\ind{\cs}\bigl\{\sop\hoch{\cs}_z \Tr\ind{\env}
  \ketbra{\varepsilon}{\varepsilon} \bigr\} \\
  &=
  \sum_\varepsilon\Tr\bigl\{
  (\sop\hoch{\cs}_z\otimes\op{1}\hoch{\env})
  \ketbra{\varepsilon}{\varepsilon} \bigr\} =
  \Tr\bigl\{
  (\sop\hoch{\cs}_z\otimes\op{1}\hoch{\env}) \bigr\},
\end{align*}
the last equality follows from the fact that the energy eigenvectors
are a complete orthonormal basis in Hilbert space.  If we now use the
basis~$\{\ket{0,m'},\ket{1,m}\}$ to calculate the trace~($m$ and~$m'$
denote the levels in band~$k$ and~$k'$, respectively), we see that the
$g$~kets~$\ket{1,m}$ yield~$1$'s, while the $g'$~kets~$\ket{0,m'}$
yield~$-1$'s and the total trace equals~$g-g'$.

\section{Derivation of eq.~(\ref{eq:5})}
\label{sec:app-a}

If we introduce the basis~$\{\ket{k{:}m}\}$ ($1<m\le g_k$) for the
band~$k$ and the basis~$\{\ket{k'{:}m}\}$ for the band~$k'$, we can
write the state of the total system as~($g=g_k$,~$g'=g_{k'}$)
\begin{equation*}
  \ket{\psi} = \sum_{m=1}^{g'} \psi_{0m}\ket{0}\otimes\ket{k'{:}m} +
  \sum_{m=1}^g \psi_{1m}\ket{1}\otimes\ket{k{:}m}\!.
\end{equation*}
The reduced state of the central system becomes
\begin{equation*}
  \op\rho\ind{\cs} = \Tr\ind{\env}\op\rho = 
  \sum_{m=1}^{g'} \abs{\psi_{0m}}^2 \ketbra{0}{0} +
  \sum_{m=1}^g \abs{\psi_{1m}}^2 \ketbra{1}{1}.
\end{equation*}
We are interested in the distribution of the inversion of the energy
eigenstates of certain Hamiltonians.  Since the inversion is
determined by the population of each level, we will derive the
distribution for the population of the ground
state,~$p_0=\sum_m\abs{\psi_{0m}}^2$.

For simplicity we introduce a single index~$n$ to label the amplitudes
instead of the double index~$0m$ or~$1m$.  $n$ runs from~$1$
to~$d=g+g'$ (the reduced Hilbert space dimension).  In this
notation~$p_0=\sum_{n=1}^{g'}\abs{\psi_n}^2$.

We now split the amplitudes into real and imaginary
part,~$\psi_n=x_k+\iu{}x_{k+1}$, where~$1\le{}k\le{}2g'$ corresponds
to~$\ketbra{0}{0}$ and~$2g'+1\le{}k\le{}2d$ corresponds
to~$\ketbra{1}{1}$,
therefore~$p_0=\sum_{k=1}^{2g'}x_k$.  The
combined probability density of the first~$2g'$ amplitudes~$x_k$
for eigenvectors of random matrices from the GUE is given
by~\cite{haake:quant_chaos}
\begin{equation*}
  P\ind{a}(x_1,\ldots,x_{2g'}) = \pi^{-1/2}
  \frac{\varGamma(d)}{\varGamma(d-g')}
  \biggl( 1 - \sum_{k=1}^{2g'} x_k^2 \biggr)^{d-g'-1}.
\end{equation*}
The desired probability density for the population of the ground
state~$p_0$ is given by
\begin{equation*}
  P\ind{p}(p_0)= \int\dd^{2g'}\!x\,
  P\ind{a}(x_1,\ldots,x_{2g'}) \, \delta\biggl(
  p_0 - \sum_{k=1}^{2g'}x_k^2 \biggr),
\end{equation*}
integrating over the unit sphere in~$2g'$-dimensional space.  The
integral yields
\begin{equation*}
  P\ind{p}(p_0) = \frac{\varGamma(d)}{\varGamma(d-g')\varGamma(g')}\,
  p_0^{g'-1} (1-p_0)^{d-g'-1}.
\end{equation*}
Transforming to~$\lambda=1-2p_0$ finally gives the probability
density~\eqref{eq:5},
\begin{equation}
  \label{eq:7}
  P(\lambda) = \frac{1}{2^{d-1}}
  \frac{\varGamma(d)}{\varGamma(d-g')\varGamma(g')}
  (1-\lambda)^{g'-1} (1+\lambda)^{d-g'-1}.
  \tag{\ref{eq:5}'}
\end{equation}

\providecommand{\sortnoop}[1]{} \providecommand{\printfirst}[2]{#1}
  \providecommand{\oneletter}[1]{#1} \providecommand{\swapargs}[2]{#2#1}
  \providecommand{\arXiv}[1]{#1} \hyphenation{}

\end{document}